\documentclass[aps,prb,twocolumn,amsmath,amssymb,showpacs,superscriptaddress]{revtex4-1}

\usepackage{graphicx}
\usepackage{amsmath,amssymb,amsfonts}
\usepackage{textcomp}
\usepackage{hyperref}
\usepackage{gensymb}
\usepackage{verbatim}
\usepackage{amsmath}
\hypersetup{breaklinks=true,colorlinks=true,urlcolor=black}
\usepackage{color}
\usepackage{soul}
\DeclareGraphicsExtensions{.jpg,.pdf,.png,.eps}

\begin{document}
\title{Pressure-temperature phase diagram of EuRbFe$_4$As$_4$ superconductor}
\author{Li Xiang}
\author{Sergey L. Bud'ko}
\affiliation{Ames Laboratory, Iowa State University, Ames, Iowa 50011, USA}
\affiliation{Department of Physics and Astronomy, Iowa State University, Ames, Iowa 50011, USA}
\email[]{ives@iastate.edu}
\author{Jin-Ke Bao}
\author{Duck Young Chung}
\author{Mercouri G. Kanatzidis}
\affiliation{Materials Science Division, Argonne National Laboratory, Argonne, Illinois 60439, USA}
\author{Paul C. Canfield}
\affiliation{Ames Laboratory, Iowa State University, Ames, Iowa 50011, USA}
\affiliation{Department of Physics and Astronomy, Iowa State University, Ames, Iowa 50011, USA}
\email[]{canfield@ameslab.gov}

\date{\today}

\begin{abstract}
	The pressure dependencies of the magnetic and superconducting transitions, as well as that of the superconducting upper critical field are reported for single crystalline EuRbFe$_4$As$_4$. Resistance measurements were performed under hydrostatic pressures up to 6.21 GPa and in magnetic fields up to 9 T. Zero-field-cool magnetization measurements were performed under hydrostatic pressures up to 1.24 GPa under 20 mT applied field. Superconducting transition temperature, $T_\text c$, up to 6.21 GPa and magnetic transition temperature, $T_\text M$, up to 1.24 GPa were obtained and a pressure-temperature phase diagram was constructed. Our results show that $T_\text c$ is monotonically suppressed upon increasing pressure. $T_\text M$ is linearly increased up to 1.24 GPa. For the studied pressure range, no signs of the crossing of $T_\text M$ and $T_\text c$ lines are observed. The normalized slope of the superconducting upper critical field is gradually suppressed with increasing pressure, which may be due to the continuous change of Fermi-velocity $v_F$ with pressure.
\end{abstract}

\maketitle 

\section{Introduction}
New members of the Fe-based superconductors (FeSC) family, $AeA$Fe$_4$As$_4$ ($Ae$=Ca, Sr; $A$=K, Rb, Cs), the so-called 1144-compounds were discovered by Iyo $et$ $al$ in 2016\cite{Iyo2016}. Different from a homogeneous, random substitution, as in ($Ae_{0.5}A_{0.5}$)Fe$_2$As$_2$ where $Ae/A$ share the same crystallographic site and retains the parent-compound symmetry $I4/mmm$, these new members crystallize into structural type $P4/mmm$ where $Ae$ and $A$ have their own unique crystallographic sites and form alternating layers along the $c$ axis\cite{Iyo2016,Kawashima2016}. Since discovery, the 1144-compounds have received significant attention because these stoichiometric compounds offer new, clean platforms for the study of, among other things, the relation between superconductivity and possible long-range magnetic order in the FeSC. Moreover, a new type of magnetic order, spin-vortex-crystal-order, has been realized in Co- and Ni-substituted CaKFe$_4$As$_4$, which was argued to be strongly related to its structure\cite{Meier2018}.

Among the new 1144 compounds, the Eu(Rb,Cs)Fe$_4$As$_4$ compounds have been studied intensively due to the possible coexistence of superconductivity and ferromagnetism \cite{Kawashima2016,Liu2016,Bao2018,Smylie2018,Stolyarov2018,Albedah2018,Stolyarov2018}. Polycrystalline Eu(Rb,Cs)Fe$_4$As$_4$ compounds were first discovered in 2016 and were shown to be superconductors with $T_\text c \sim$ 35 K and a magnetic transition temperature $T_\text M \sim$ 15 K\cite{Kawashima2016}. Different from the undoped EuFe$_2$As$_2$ where Eu$^{2+}$ orders antiferromagnetically\cite{Ren2008,Jeevan2008,Jiang2009a}, the magnetic transition in RbEuFe$_4$As$_4$ is suggested to be ferromagnetic which is associated with the ordering of the Eu$^{2+}$ moments perpendicular to the crystallographic $c$ axis\cite{Liu2016,Albedah2018}. Though the exact magnetic structure of EuRbFe$_4$As$_4$ has not been established so far, the possible coexistence of superconductivity and ferromagnetism makes EuRbFe$_4$As$_4$ one of the systems where the relation between these states may be studied\cite{Fertig1977,Ishikawa1977,Canfield1996,Saxena2000,Aoki2001,Pfleiderer2001,Huy2007,Jiang2009b,Nowik2011,Jiao2011,Jiao2013,Jin2013,Jin2015}.

Two substitution studies on polycrystalline EuRbFe$_4$As$_4$ were published. On one hand, Ni-substitution on the Fe-site suppresses $T_\text c$ whereas $T_\text M$ is almost unchanged\cite{Liu2017}. On the other hand, substitution of non-magnetic Ca on the Eu-site suppresses $T_\text M$ while $T_\text c$ is almost unchanged\cite{Kawashima2018}. Both of these results suggest that superconductivity and ferromagnetism are almost indenpendent of each other in this system. An optical investigation on single crystalline EuRbFe$_4$As$_4$ suggests weak interaction between superconductivity and ferromagnetism and that superconductivity is affected by the in-plane ferromagnetism mainly at domain boundaries\cite{Stolyarov2018}.

Pressure, as another commonly used tuning parameter, is considered less perturbing than substitution because it does not introduce chemical disorder into the system. A high pressure study up to $\sim$ 30 GPa on polycrystalline Eu(Rb,Cs)Fe$_4$As$_4$ shows that for both compositions, upon increasing pressure, $T_\text c$ is suppressed while $T_\text M$ is enhanced and they cross near 7 GPa\cite{Jackson2018}. In addition, half-collapsed-tetragonal (hcT) phase transition, similar to the one observed in the CaKFe$_4$As$_4$ series\cite{Kaluarachchi2017PRB,Xiang2018PRB}, is suggested to take place at $\sim$ 10 GPa for EuRbFe$_4$As$_4$ and $\sim$ 12 GPa for EuCsFe$_4$As$_4$, respectively\cite{Jackson2018}, which is roughly consistent with theoretical calculations\cite{Borisov2018}. In this high-pressure study, signatures of transitions are broad and zero resistance was never achieved below $T_\text c$ due, most likely, to the use of polycrystalline samples.

In this work, we present a pressure study on single crystalline EuRbFe$_4$As$_4$ up to 6.21 GPa. From resistance measurements up to 6.21 GPa and magnetization measurements up to 1.24 GPa, $T_\text c$ and $T_\text M$ are tracked and presented in a pressure-temperature ($p-T$) phase diagram. Our results show that $T_\text c$ is monotonically suppressed and $T_\text M$ is linearly increased. Further superconducting upper critical field analysis indicates no qualitative change of Fermi surface within the studied pressure range.

\section{Experimental details}
High-quality single crystals of EuRbFe$_4$As$_4$ with sharp superconducting transitions at ambient pressure (see Figs. \ref{figure1_RT} (c) (d) and Fig. \ref{figure5_MT} (b) below) were grown as described in Ref. \onlinecite{Bao2018}. The $ab$-in-plane ac resistance measurements under pressure were performed in a Quantum Design Physical Property Measurement System (PPMS) using a 1 mA excitation with frequency of 17 Hz, on cooling rate of 0.25 K/min. A standard, linear four-contact configuration was used. Contacts were made by soldering 25 $\mu$m Pt wires to the samples using a Sn:Pb-60:40 alloy. The magnetic field was applied along the $c$ axis. A modified Bridgman Anvil Cell (mBAC)\cite{Colombier2007} was used to apply pressure up to 6.21 GPa. Pressure values at low temperature were inferred from the $T_\textrm{c}(p)$ of lead\cite{Bireckoven1988}. Hydrostatic conditions were achieved by using a 1:1 mixture of iso-pentane:n-pentane as the pressure medium for the mBAC, which solidifies at $\sim$ 6.5 GPa at room temperature\cite{Torikachvili2015}. Low-field (20 mT) dc magnetization measurements under pressure were performed in a Quantum Design Magnetic Property Measurement System (MPMS-3) SQUID magnetometer. A commercially-available HDM Be-Cu piston-cylinder pressure cell\cite{HDM} was used to apply pressures up to 1.24 GPa. Daphne oil 7373 was used as a pressure medium, which solidifies at ∼2.2 GPa at room temperature\cite{Yokogawa2007}, ensuring hydrostatic conditions. Superconducting Sn was used as a low-temperature pressure gauge\cite{Eiling1981}. Two samples, $\#$1 and $\#$2, were used for separate resistivity runs in the mBAC whereas the sample used for the magnetization data was sample $\#$3.

\section{Results and discussions}
	
\begin{figure}
	\includegraphics[width=8.6cm]{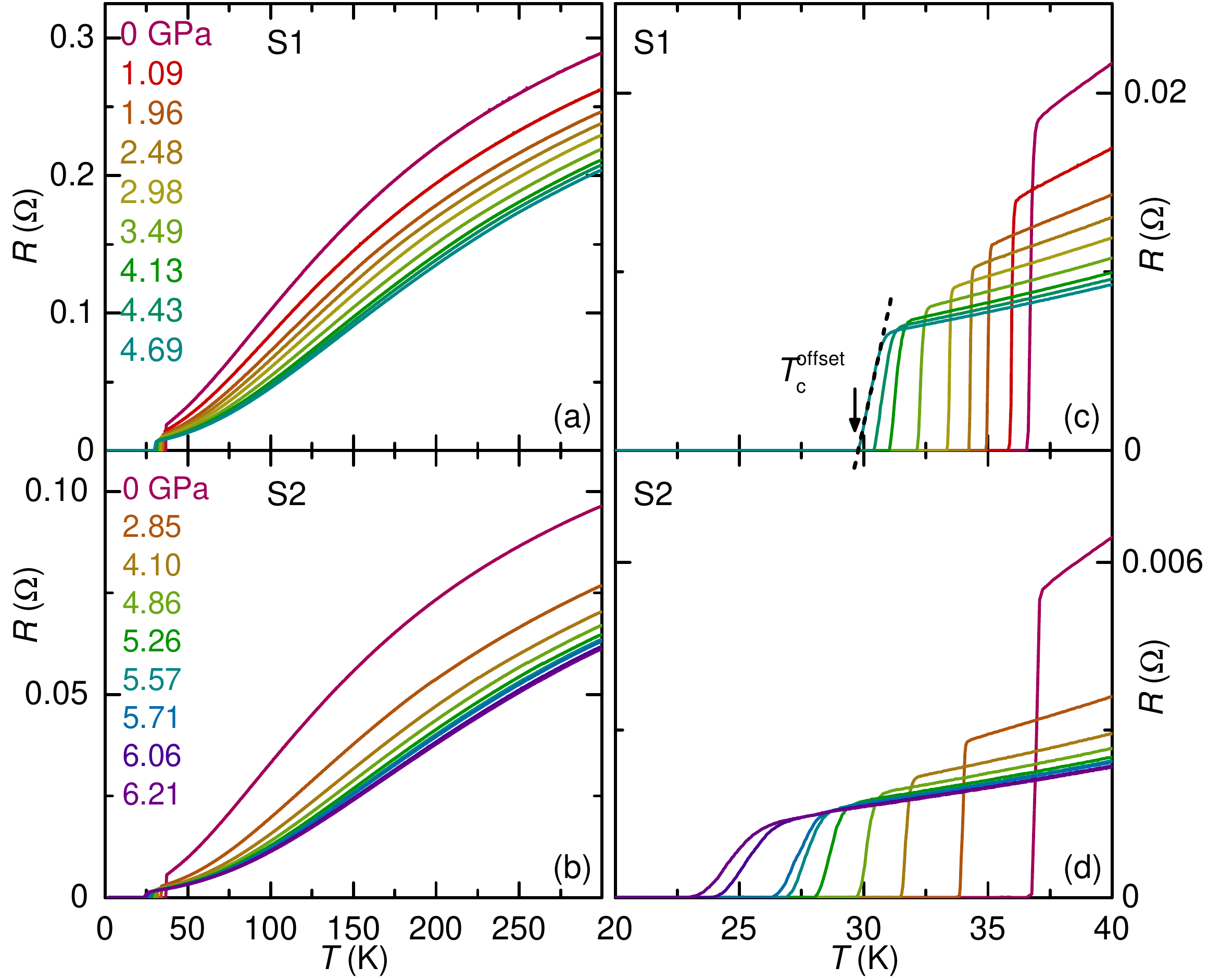}%
	\caption{(a) (b) Evolution of the in-plane resistance with hydrostatic pressures up to 6.21 GPa measured in a mBAC for EuRbFe$_4$As$_4$ sample \#1 and sample \#2, respectively. (c) (d) Blowups of the low temperature region showing the superconducting transition. Criterion for $T_\textrm{c}^\textrm{offset}$ is indicated by arrow in (c).
		\label{figure1_RT}}
\end{figure}

\begin{figure}
	\includegraphics[width=8.6cm]{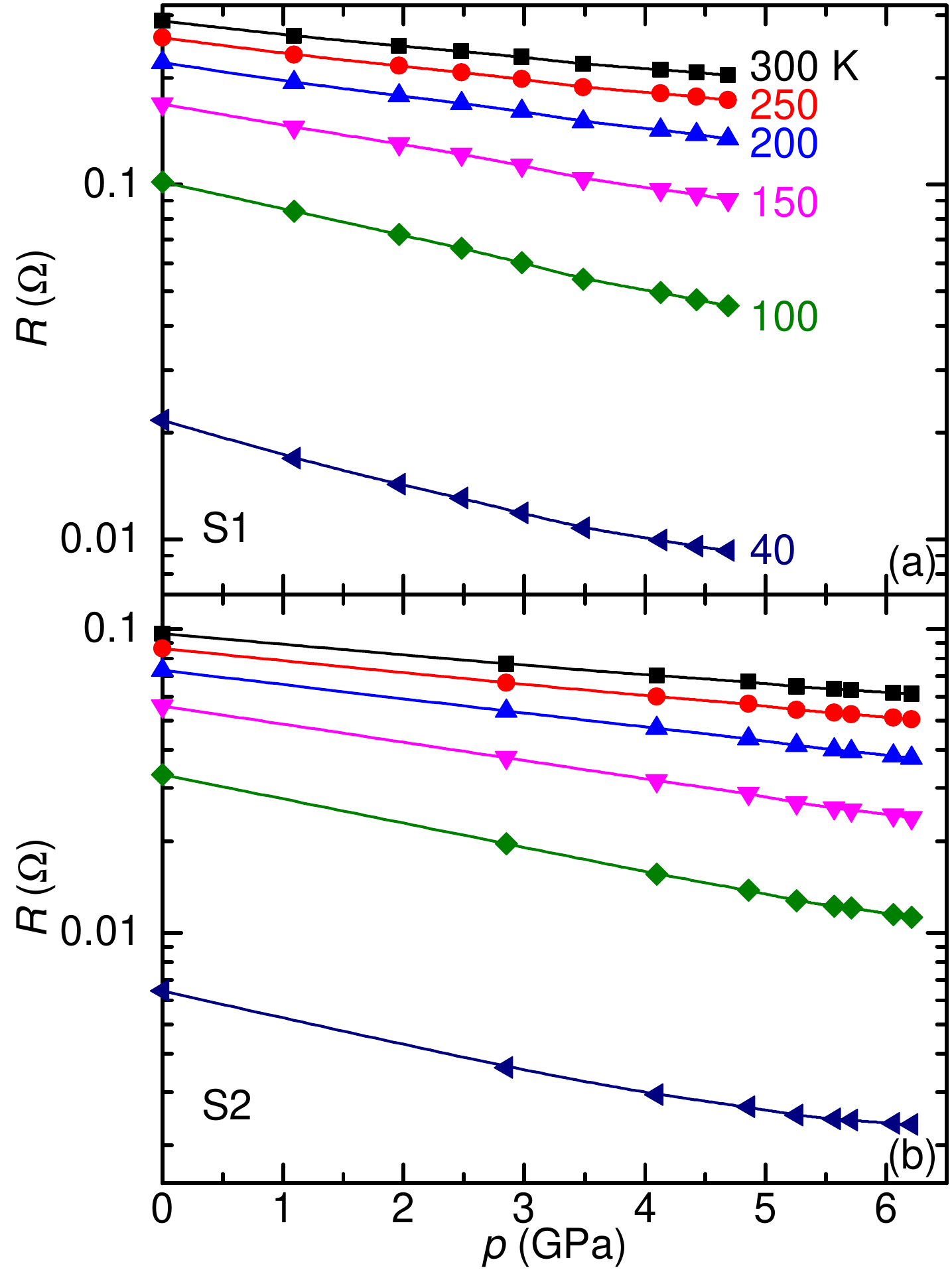}%
	\caption{Pressure dependence of resistance $R(p)$ at fixed temperatures for EuRbFe$_4$As$_4$ sample \#1 (a) and sample \#2 (b).
		\label{R_p}}
\end{figure}

\begin{figure}
	\includegraphics[width=8.6cm]{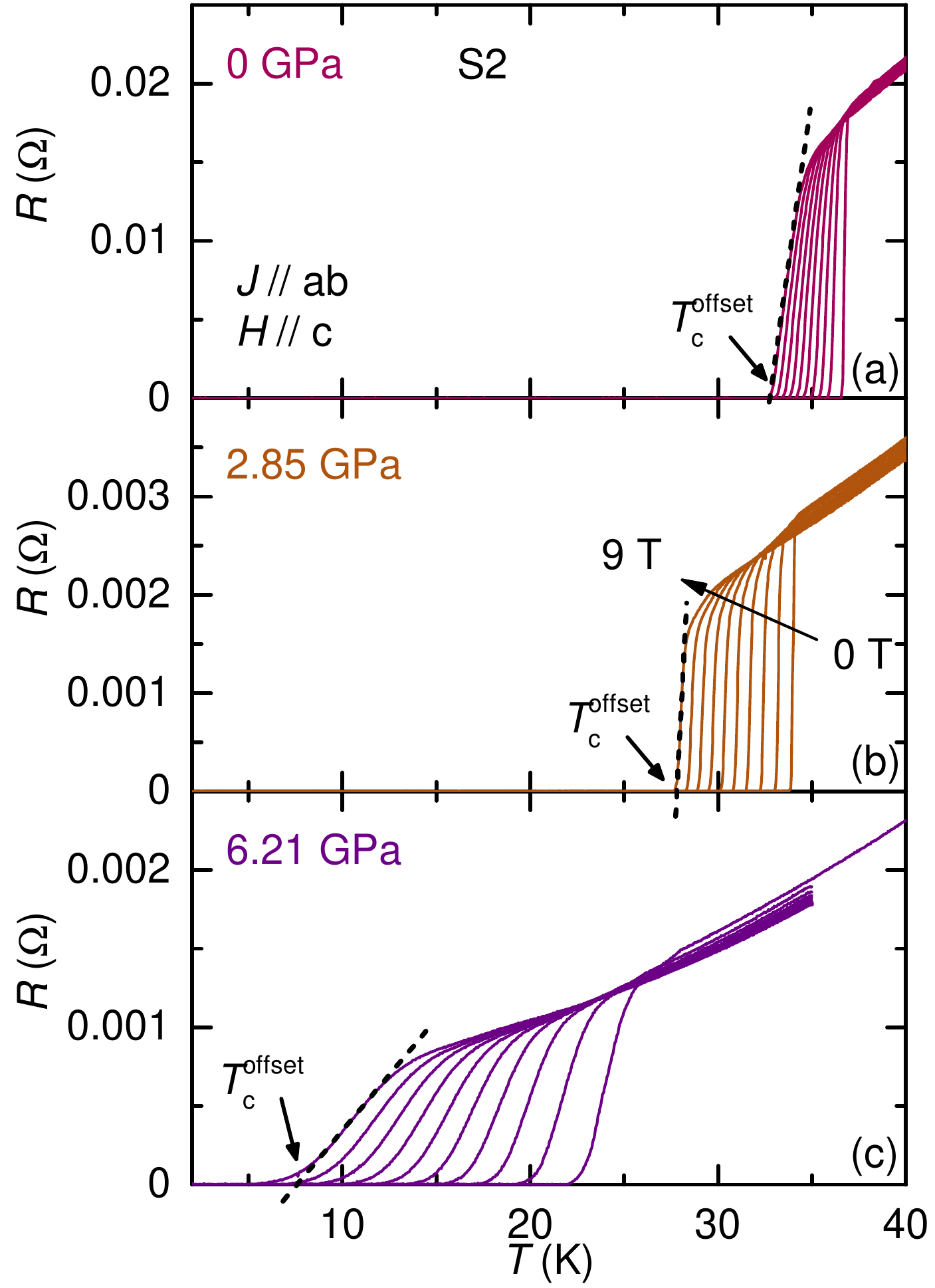}%
	\caption{Temperature dependence of resistance under magnetic field up to 9 T for selective pressures for sample \#2. Criteria for $T_\textrm{c}^\textrm{offset}$ under magnetic fields are indicated by arrows. Current was applied in-plane and magnetic field was applied along $c$ axis.
		\label{figure2_RTH}}
\end{figure}

\begin{figure}
	\includegraphics[width=8.6cm]{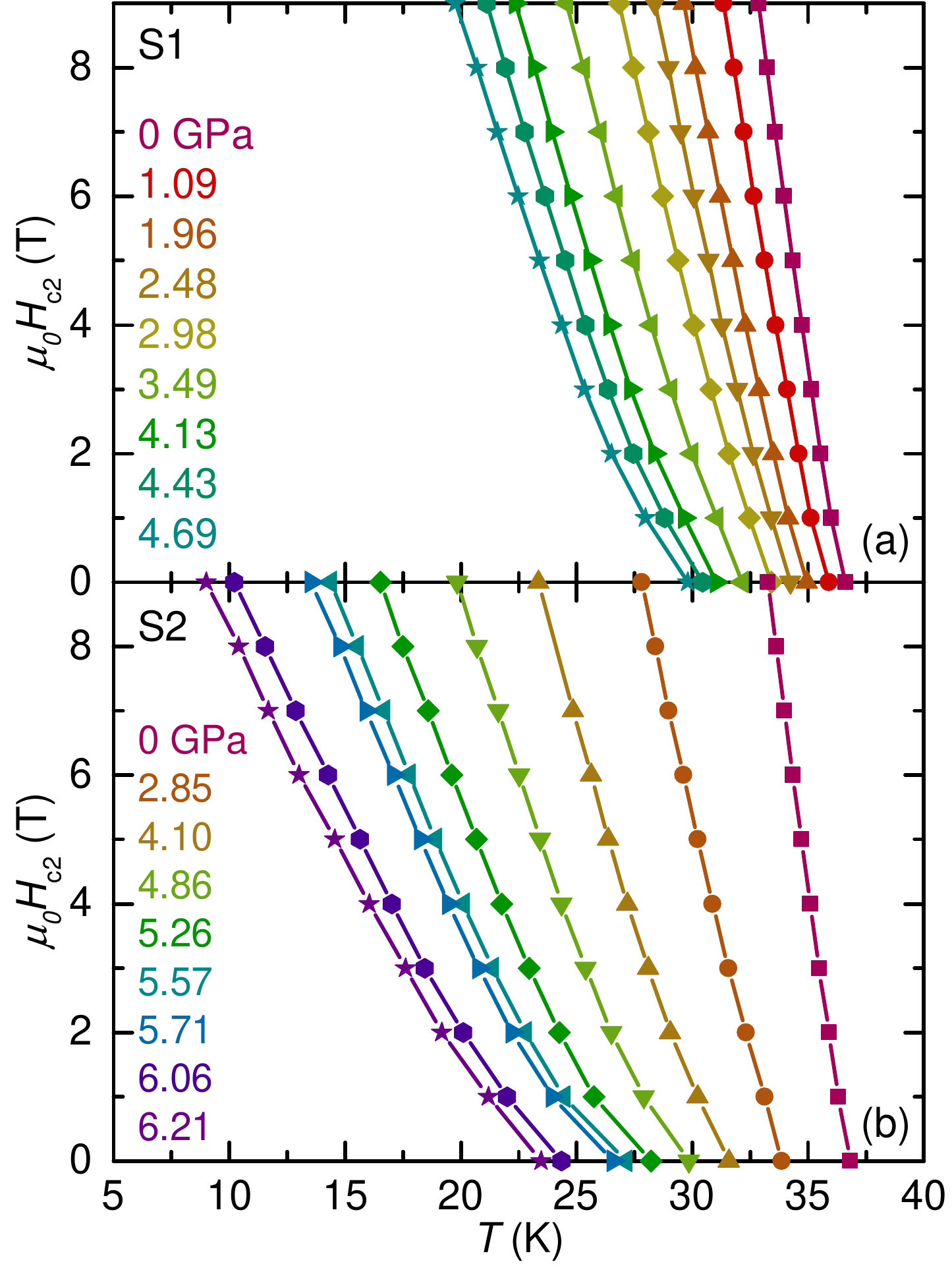}%
	\caption{Temperature dependence of the upper superconducting critical field, $H_{\textrm c2}(T)$, under selected pressures for (a) sample \#1 and (b) sample \#2.
		\label{figure3_Hc2}}
\end{figure}

%\begin{figure}
%	\includegraphics[width=8.6cm]{figure4_MR_40K}%
%	\caption{Evolution of the MR measured at $T=$ 40 K with pressures up to 5.57 GPa for (a) sample\#3 and (b) sample\#2, respectively. Magnetic fields were applied along the $c$ axis.
%		\label{figure4_MR_40K}}
%\end{figure}

Figures \ref{figure1_RT} (a) and (b) present the pressure dependence of the temperature-dependent resistance for EuRbFe$_4$As$_4$. Two samples, sample \#1 and sample \#2, were measured in the mBAC for pressures up to 4.69 GPa or 6.21 GPa. For both samples, resistance decreases upon increasing pressure. At ambient pressure for $T \sim$ 35 K, a superconducting transition was observed and zero resistance was achieved for both samples. Below $T_\text c$, no features associated with the magnetic transition $T_\text M$ are observed in the $R(T)$ curves down to 1.8 K. Figs. \ref{figure1_RT} (c) and (d) show blowups of the low-temperature resistance. For both samples, the superconducting transition at ambient pressure is very sharp, demonstrating good homogeneity of the single crystals. As shown in the figures, upon increasing pressure, $T_\text c$ monotonically decreases in the studied pressure range. A gradually broadening of the superconducting transition was also observed in both samples. Similar behavior has been observed in many other superconductors that are measured in the mBAC cell and is likely due to the pressure inhomogeneity when high loads are applied.

To better visualize the pressure evolution of resistance, we present in Fig. \ref{R_p} the pressure dependent resistance $R(p)$ at fixed temperatures. As shown in the figure, different from the CaKFe$_4$As$_4$ series\cite{Kaluarachchi2017PRB,Xiang2018PRB}, resistance of EuRbFe$_4$As$_4$ at various temperatures shows a smooth decrease as a function of pressure without any obvious anomalies. This implies the absence of structural transition up to 6.21 GPa, which is consistent with the results in Ref. \onlinecite{Jackson2018} and predictions in Ref. \onlinecite{Borisov2018} where the hcT phase transtion is suggested to take place at $\sim$ 10 GPa. The total suppression of resistance at 40 K under pressure, $\sim 55\%$ up to 4 GPa and $\sim 65\%$ up to 6.21 GPa, is rather large compared with the CaKFe$_4$As$_4$ series, where the suppression at 40 K is $30\%$ - $40\%$ up to 4 GPa, i.e., before hcT happens\cite{Kaluarachchi2017PRB,Xiang2018PRB}. Another indication that a potential hcT phase transition has not been reached is the fact that the superconducting transitions shown in Figs. \ref{figure1_RT} and \ref{figure2_RTH} remain robust over our pressure range. Both CaKFe$_4$As$_4$ series\cite{Kaluarachchi2017PRB,Xiang2018PRB} as well as Co-substituted CaFe$_2$As$_2$\cite{Ran2012,Gati2012} show loss of bulk superconductivity at the collapsed-tetragonal or lowest hcT transitions.

Temperature dependent resistance under magnetic fields up to 9 T applied along the $c$-axis was studied and the results are presented in Fig. \ref{figure2_RTH} for selected pressures for sample \#2. As shown in the figure, below $T_\text c$, no features associated with the magnetic transition $T_\text M$ are observed and zero resistance persists down to 1.8 K with fields up to 9 T under all pressures. For temperatures above the superconducting transition, a decrease of resistance under applied magnetic field is observed. The upper superconducting critical field, $H_\text{c2}$, can be obtained from Fig. \ref{figure2_RTH} using the offset criteria defined in Figs. \ref{figure1_RT}-\ref{figure2_RTH}. The temperature dependence of $H_\text{c2}$ at various pressures for sample \#1 and sample \#2 is presented in Fig. \ref{figure3_Hc2}. For both samples, $H_\text{c2}$ is systematically suppressed by increasing pressure. $H_\text{c2}$ is linear in temperature except for magnetic fields below 2 T, the bending of $H_\text{c2}(T)$ curves are more obvious at higher pressures. The curvature at low fields has been observed in other FeSC\cite{Colombier2009,Colombier2010,Kaluarachchi2016,Xiang2017PRB,Xiang2018PRB} and can be explained by the multi-bands nature of superconductivity\cite{Kogan2012}, which is likely the case of EuRbFe$_4$As$_4$\cite{Stolyarov2018}.

%We further measured the magnetoresistance of EuRbFe$_4$As$_4$ at $T=$ 40 K under various pressures. The field dependence of the MR with field applied along the $c$ axis, defined as MR = $\frac{R(H)-R(0)}{R(0)}\times 100\%$, at various pressures is presented in Fig. \ref{figure4_MR_40K}. As shown in the figure, the resistance decreases with increasing magnetic fields, which is a typical behavior for metals at paramagnetic state\cite{?????}. Resistance decreases due to decreasing of spin disorder scattering when external magnetic field is applied.???? As pressure is increased, the drop of MR value is enhanced. As shown in the figure, MR value drops $\sim$ 3\% at ambient pressure, it drops $\sim$ 7\% at 6.21 GPa. This might be due to ???????

\begin{figure}
	\includegraphics[width=8.6cm]{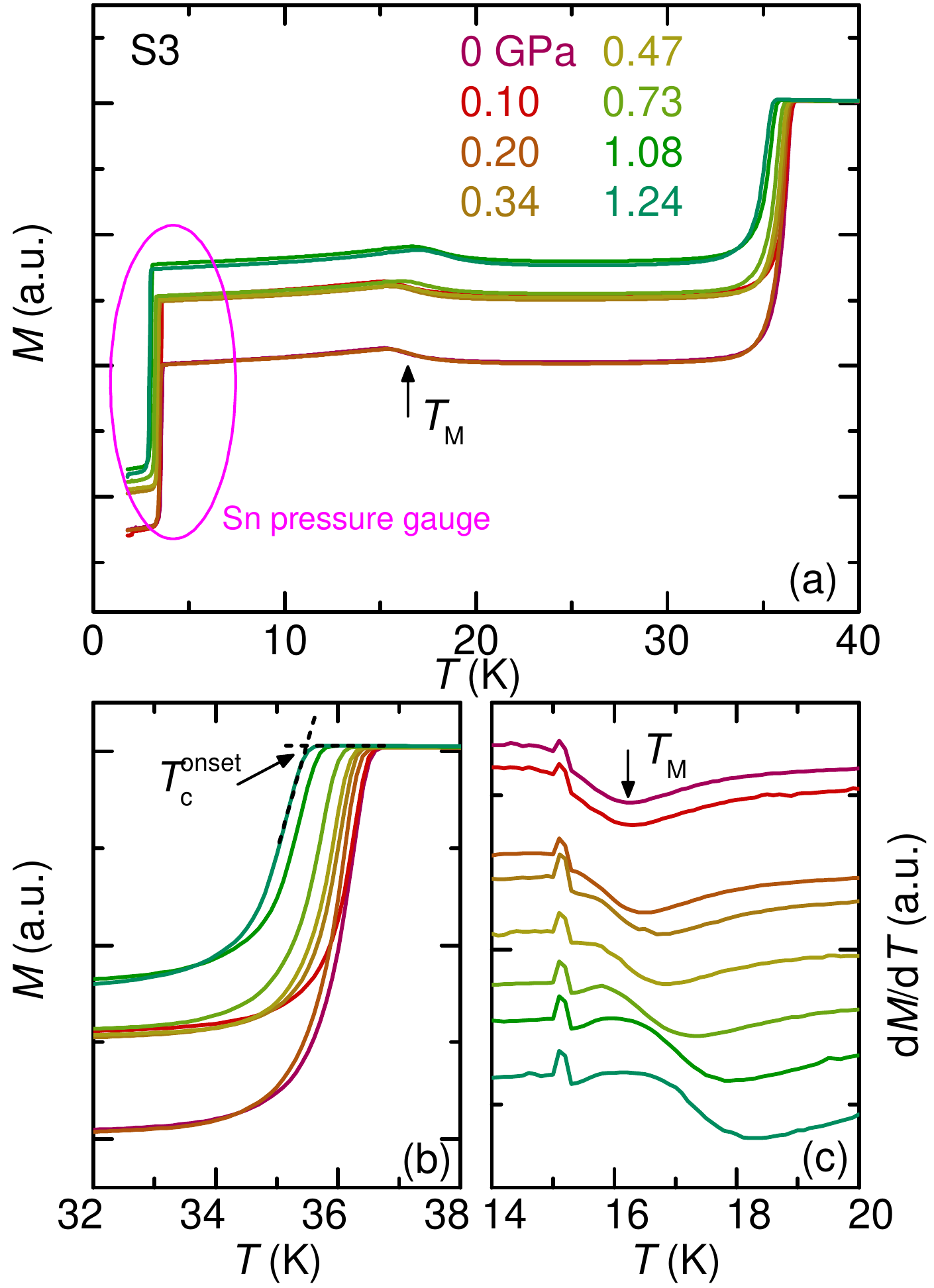}%
	\caption{(a) Evolution of the zero-field-cool (ZFC) magnetization $M(T)$ with hydrostatic pressures up to 1.24 GPa under 20 mT applied field. Superconducting transition of Sn is used to determined the low temperature pressure, as indicated by the pink circle. (b) Blow up of the superconducting transition region for EuRbFe$_4$As$_4$. Criterion for $T_\text c^\text{onset}$ is indicated by arrow. (c) Temperature derivative of the magnetization, $dM/dT$, showing the evolution of the magnetic transition $T_\textrm{M}$. Criterion is indicated by arrow. The small feature just above 15 K is an artifact caused by the combination of small temperature steps and details of the temperature control in MPMS-3.
		\label{figure5_MT}}
\end{figure}

\begin{figure}
	\includegraphics[width=8.6cm]{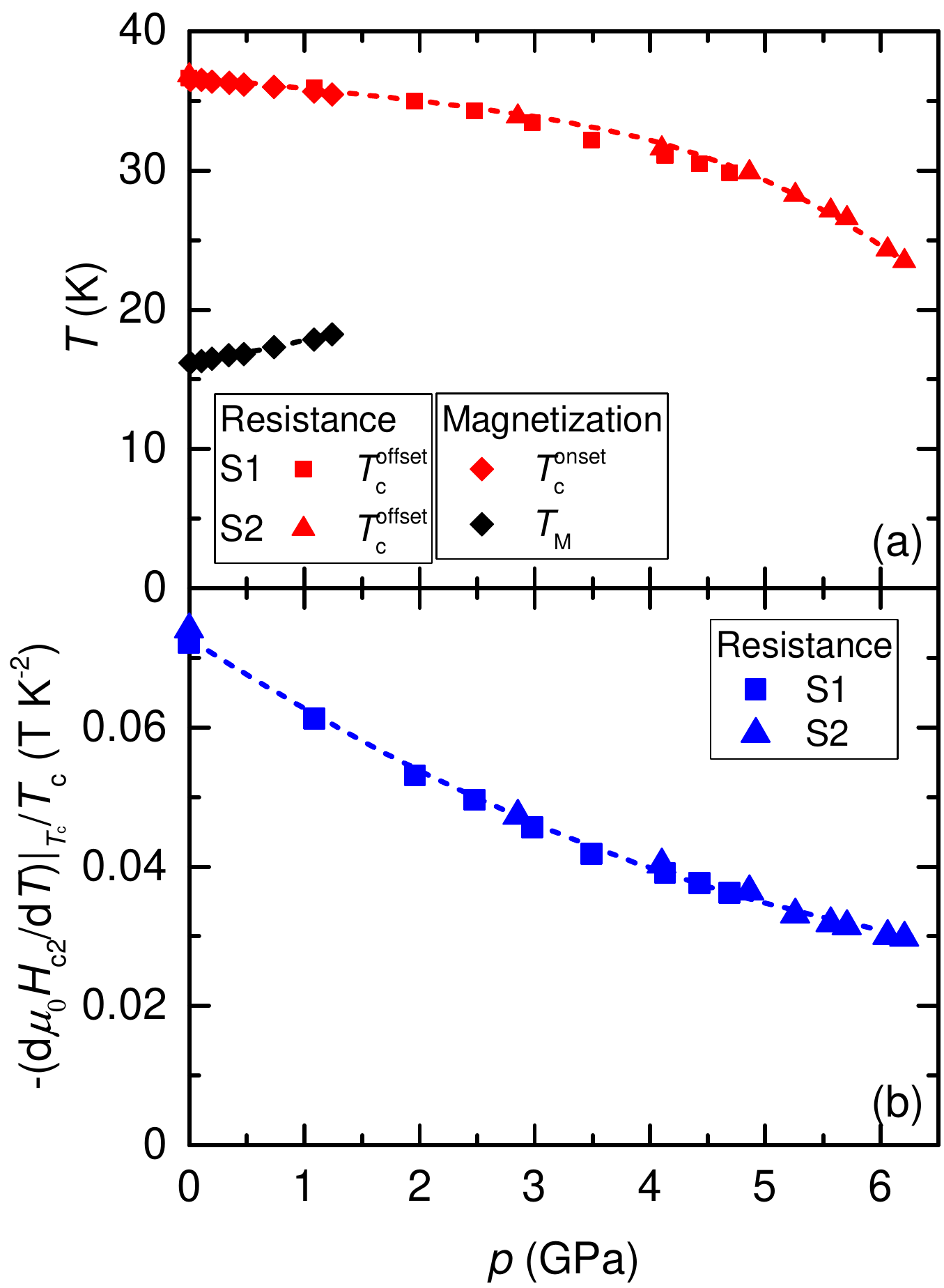}%
	\caption{(a) Pressure-temperature phase diagram of EuRbFe$_4$As$_4$, as determined from resistance and magnetization measurements. Red and black symbols represent the superconducting $T_\textrm{c}^\textrm{offset}$ and magnetic $T_\textrm{M}$ phase transitions. (b) Pressure dependence of the normalized upper critical field slope -(1/$T_\textrm c$)($d\mu_oH_{\textrm c2}$/$dT$)$|_{T_\textrm c}$. The squares and triangles are data obtained from resistance measurement for sample \#1 and sample \#2, respectively. The diamonds are data obtained from magnetization measurement. Dashed lines are guides to the eye.
		\label{figure6_phasediagram}}
\end{figure}

To study the evolution of the magnetic transition with pressure, we present, in Fig. \ref{figure5_MT}, the dependence of the zero-field-cool magnetization $M(T)$ data. During the measurements, pressure was increased up to 1.24 GPa under 20 mT applied magnetic field. As shown in Fig. \ref{figure5_MT} (a), the superconducting transition of EuRbFe$_4$As$_4$ is recognized as onset of diamagnetism at $T \sim$ 35 K. Another kink-like anomaly is observed at $T \sim$ 16 K. We associated this anomaly with the magnetic transition $T_\text M$. Pressure values at low temperature were inferred from the superconducting transition of Sn which also shown up in the data set at $T \sim$ 3.7 K, i.e., way below $T_\text c$ and $T_\text M$ of EuRbFe$_4$As$_4$ (as indicated inside the pink circle in the figure). Fig. \ref{figure5_MT} (b) shows the blowup of the superconducting transition region of EuRbFe$_4$As$_4$, demonstrating that $T_\text c$ is suppressed as pressure is increased. To determine the magnetic transition temperature $T_\text M$, temperature derivative of the magnetization, $dM/dT$, was calculated and presented in Fig. \ref{figure5_MT} (c). The temperature corresponding to the minimum in $dM/dT$ was taken as $T_\text M$, as indicated in the figure. It is clearly seen that $T_\text M$ is increased upon increasing pressure.

We summarize the $T_\text c$ and $T_\text M$ values inferred from both resistance and magnetization measurements in the pressure-temperature ($p-T$) phase diagram shown in Fig. \ref{figure6_phasediagram} (a). To be consistent, $T_\textrm{c}^\textrm{offset}$ determined from resistance measurements (Fig. \ref{figure1_RT} (c)) and $T_\text c^\text{onset}$ determined from magnetization measurements (Fig. \ref{figure5_MT} (b)) were used and they match with each very well. As shown in Fig. \ref{figure6_phasediagram} (a), $T_\text c$ of EuRbFe$_4$As$_4$ is monotonically suppressed upon increasing pressure up to 6.21 GPa. Starting with $T_\text c=$ 36.6 K at ambient pressure, $T_\text c$ is suppressed to 23.5 K at 6.21 GPa. In terms of magnetic transition $T_\text M$, it is linearly increased from 16.2 K at ambient pressure to 18.2 K at 1.24 GPa, with the rate of $dT_\text M/dp$ = 1.64 K/GPa. To better understand the superconducting properties of EuRbFe$_4$As$_4$, we further analyze the superconducting upper critical field\cite{Taufour2014,Kaluarachchi2016,Xiang2017PRB,Xiang2018PRB}. Generally speaking, the slope of the upper critical field normalized by $T_\textrm{c}$, is related to the Fermi velocity and superconducting gap of the system\cite{Kogan2012}. In the clean limit, for a single-band,
\begin{equation}
-(1/T_\textrm c)(d\mu_oH_{\textrm c2}/dT)|_{T_\textrm c} \propto 1/v_F^2,
\label{eq:Hc2}
\end{equation}
where $v_F$ is the Fermi velocity. Even though the superconductivity in EuRbFe$_4$As$_4$ compounds is likely to be multiband, Eq. \ref{eq:Hc2} can give qualitative insight into changes induced by pressure. As shown in Fig. \ref{figure6_phasediagram} (b), the normalized slope of the upper critical field $-(1/T_\textrm c)(d\mu_oH_{\textrm c2}/dT)|_{T_\textrm c}$ (the slope $d\mu_oH_{\textrm c2}/dT$ is obtained by linearly fitting the data above 2 T in Fig. \ref{figure3_Hc2}) is gradually suppressed by a factor of $\sim$ 2.5 upon increasing pressure up to 6.21 GPa. No features in the normalized slope that could be associated with band structure change or Lifshitz transition, like the cases in many other Fe-based superconductors\cite{Taufour2014,Kaluarachchi2016,Xiang2017PRB,Xiang2018PRB}, are observed over the studied pressure range. Furthermore, the $R(p)$ curve at 40 K (Fig. \ref{R_p} (b)), a temperature that is close to $T_\text c$ but still above $T_\text c$ and $T_\text M$, implies that resistivity is suppressed by a factor of $\sim$ 2.7 as well. In a simple argument\cite{Kasap2006},
\begin{equation}
\rho \propto 1/(g_{\epsilon_F}\tau v_F^2)
\label{eq:rho}
\end{equation}
where $g_{\epsilon_F}$ is density of states at the Fermi level and $\tau$ is the scattering time of these Fermi electrons. Eq. \ref{eq:Hc2} and \ref{eq:rho}, combined together, suggest that the decrease of both resistivity and $-(1/T_\textrm c)(d\mu_oH_{\textrm c2}/dT)|_{T_\textrm c}$ with pressure can be explained by pressure induced increase of the Fermi velocity.
\begin{figure}
	\includegraphics[width=8.6cm]{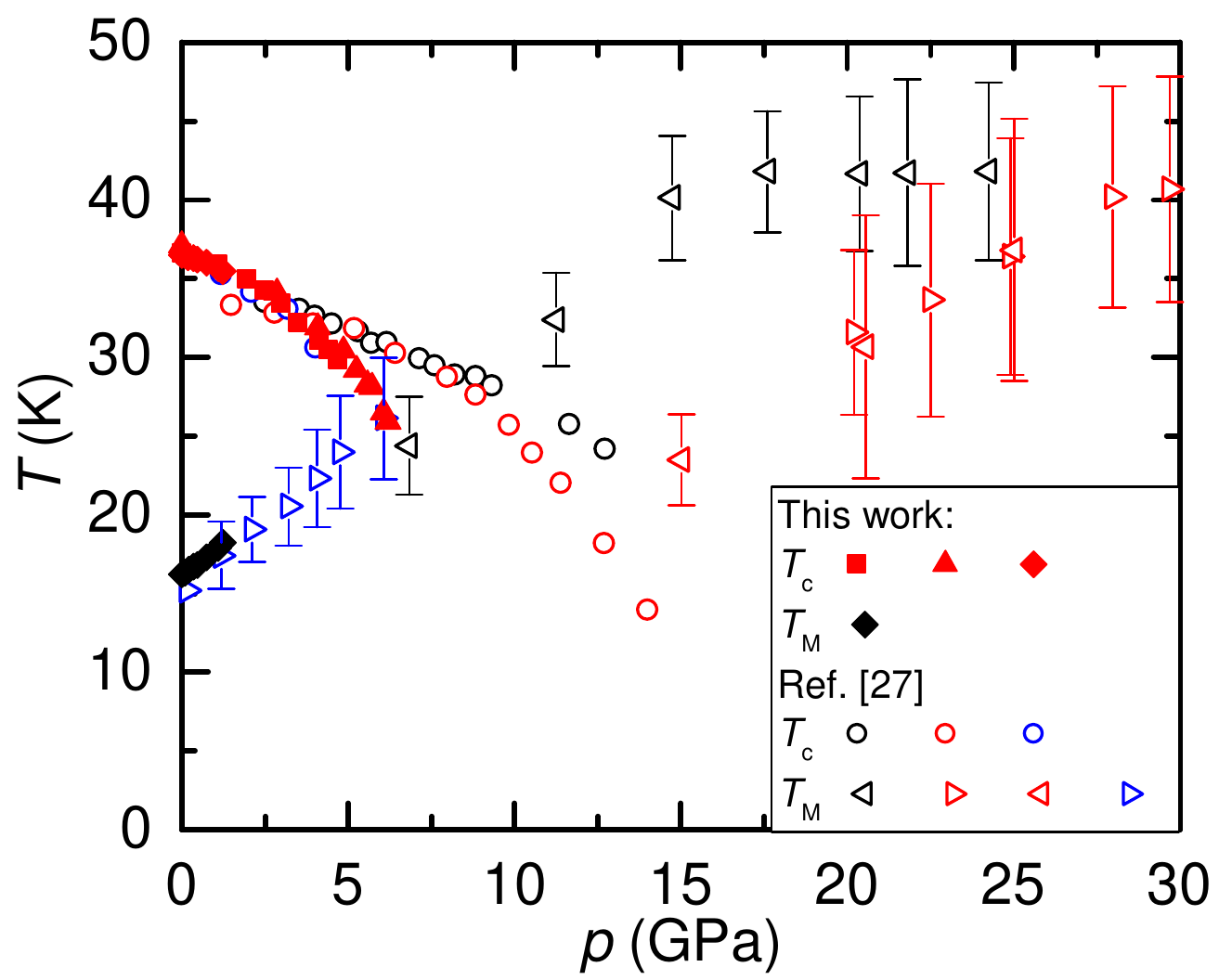}%
	\caption{Pressure-temperature phase diagram of EuRbFe$_4$As$_4$ up to $\sim$ 30 GPa, including data from Ref. \onlinecite{Jackson2018} (open symbols). Open circles corresponds to the onset of the superconducting transition measured via resistivity or magnetic susceptibility. Open triangles corresponds to the magnetic transition determined from magnetic susceptibility or feature in d$\rho$/d$T$.
		\label{phasediagram_combine}}
\end{figure}

Data from this study, on single crystalline samples, and from the study on polycrystalline samples in Ref. \onlinecite{Jackson2018} are plotted together and presented in the combined $p-T$ phase diagram in Fig. \ref{phasediagram_combine}. As shown in the figure, $T_\text c$ from this study (determined by the offset of the transition via resistance measurement or onset of diamagnetism) matches very well with the $T_ \text c$ determined by the onset of diamagnetism in Ref. \onlinecite{Jackson2018}. $T_\text M$ data also match with each other over the studied pressure range.

The extrapolation of our $T_\text M(p)$ line in Fig. \ref{figure6_phasediagram}(a) as well as the data in Fig. \ref{phasediagram_combine} suggest that $T_\text c(p)$ and $T_\text M(p)$ should cross near 6 GPa. On one hand, the suppression of $T_\text c$ with pressure gets stronger when pressure is increased, which might be related to the fact that $T_\text c(p)$ and $T_\text M(p)$ are getting closer at higher pressures.  On the other hand, neither our pressure dependent $T_\text c$ nor $-(1/T_\textrm c)(d\mu_oH_{\textrm c2}/dT)|_{T_\textrm c}$ data show any clear signature potentially associated with $T_\text c(p)$ and $T_\text M(p)$ crossing. Either they cross at a pressure higher than 6.21 GPa or their crossing does not have qualitative effect on $T_\text c(p)$ or $H_{c2}(T,p)$.
%From the combined $p-T$ phase diagram, it is shown that $T_\text c$ and $T_\text M$ are getting close to each other and almost crossing up to 6.21 GPa (solid red triangles and open blue triangles). However, as mentioned before, no significant features are observed in the normalized slope of the superconducting upper critical field up to 6.21 GPa (Fig. \ref{figure6_phasediagram} (b)). The change of the normalized slope may simply be caused by the change of $v_F$ with pressure. This suggests weak interaction between superconductivity and magnetism in this compound or $T_\text M$ is still lower than $T_\text c$ and there is no crossing of $T_\text M$ and $T_\text c$ lines up to 6.21 GPa.

\section{Conclusion}
In conclusion, the resistance and magnetization of single crystalline EuRbFe$_4$As$_4$ has been studied under pressure. In-plane resistance measurements under pressure up to 6.21 GPa reveal that superconducting transition $T_\text{c}$ is monotonically suppressed. Magnetization measurements under pressure up to 1.24 GPa reveal that magnetic transition $T_\text{M}$ is linearly increased. No indications of half-collapsed-tetragonal phase transition is observed up to 6.21 GPa. Further upper critical field analysis shows that the normalized slope, $-(1/T_\textrm c)(d\mu_oH_{\textrm c2}/dT)|_{T_\textrm c}$, is continuously suppressed upon increasing pressure up to 6.21 GPa, which is likely due to the continuous change of the Fermi velocity with pressure. Our results suggest that the magnetism of Eu sub-lattice does not have significant influence on the superconducting behavior of FeAs layer in EuRbFe$_4$As$_4$.

\begin{acknowledgements}
We would like to thank U. Welp and V. G. Kogan for some preliminary results and useful discussions. This work is supported by the US DOE, Basic Energy Sciences, Materials Science and Engineering Division under contract No. DE-AC02-07CH11358. L. X. was supported, in part, by the W. M. Keck Foundation. The synthesis and crystal growth were performed at Argonne National Laboratory which is supported by the U.S. Department of Energy, Office of Science, Basic Energy Sciences under Contract No. DE-AC02-06CH11357.
\end{acknowledgements}

\clearpage

\bibliographystyle{apsrev4-1}
%\bibliography{MyRef}
%

\end{document}